\pdfoutput=1
\documentclass[10pt,twocolumn]{article}
\usepackage{styles/ieee-style}

\author{Mathias Gottschlag, Yussuf Khalil, Frank Bellosa}
\affiliation{Operating Systems Group\\ Karlsruhe Institute of Technology}
\email{os@itec.kit.edu}

\usepackage{styles/ka-style}
\usepackage{cite,xspace,ifthen,graphicx,listings}
\usepackage{styles/ka-style}

\usepackage[hyphens]{url}
\usepackage{color}
\usepackage{graphicx}
\usepackage{pgfplots}
\usepackage{comment}
\usepackage{csquotes}
\usepackage{subcaption}
\usepackage{siunitx}
\usepackage{todonotes}
\usepackage{multirow}
\usepackage{filecontents}

\pgfplotsset{compat = 1.14}
\usetikzlibrary{arrows}
\usetikzlibrary{arrows.meta}
\usetikzlibrary{calc}
\usetikzlibrary{patterns}
\usetikzlibrary{fit}

\pgfplotsset{select coords between index/.style 2 args={
    x filter/.code={
        \ifnum\coordindex<#1\fi
        \ifnum\coordindex>#2\fi
    }
}}

\usepackage[
   pdfauthor={Mathias Gottschlag, Yussuf Khalil, Frank Bellosa},
   pdftitle={Technical Report: Dim Silicon and the Case for Improved DVFS Policies},
   pdfkeywords={AVX-512, AVX2, dark silicon, DVFS}
]{hyperref}

\definecolor{kitgreenexcl}{cmyk}{1.0,  0.0,  0.6, 0.0}
\definecolor{kitblue}     {cmyk}{0.8,  0.5,  0.0, 0.0}
\definecolor{kitgreen}    {cmyk}{0.6,  0.0,  1.0, 0.0}
\definecolor{kityellow}   {cmyk}{0.0,  0.05, 1.0, 0.0}
\definecolor{kitorange}   {cmyk}{0.0,  0.45, 1.0, 0.0}
\definecolor{kitbrown}    {cmyk}{0.35, 0.5,  1.0, 0.0}
\definecolor{kitred}      {cmyk}{0.25, 1.0,  1.0, 0.0}
\definecolor{kitpurple}   {cmyk}{0.25, 1.0,  0.0, 0.0}
\definecolor{kitcyan}     {cmyk}{0.9,  0.05, 0.0, 0.0}

\tikzset{
	hatch distance/.store in=\hatchdistance,
	hatch distance=10pt,
	hatch thickness/.store in=\hatchthickness,
	hatch thickness=2pt
}

\makeatletter
\pgfdeclarepatternformonly[\hatchdistance,\hatchthickness]{flexible hatch}
{\pgfqpoint{0pt}{0pt}}
{\pgfqpoint{\hatchdistance}{\hatchdistance}}
{\pgfpoint{\hatchdistance-1pt}{\hatchdistance-1pt}}%
{
	\pgfsetcolor{\tikz@pattern@color}
	\pgfsetlinewidth{\hatchthickness}
	\pgfpathmoveto{\pgfqpoint{0pt}{0pt}}
	\pgfpathlineto{\pgfqpoint{\hatchdistance}{\hatchdistance}}
	\pgfusepath{stroke}
}
\makeatother

\begin{document}

\title{Technical Report:\\Dim Silicon and the Case for Improved DVFS Policies}

\newcommand{\code}[1]{{\tt \small{#1}}}

\maketitle
\draftfooter

\begin{abstract}
Due to thermal and power supply limits, modern Intel CPUs reduce their frequency when AVX2 and AVX-512 instructions are executed.
As the CPUs wait for \SI{670}{\micro\second} before increasing the frequency again, the performance of some heterogeneous workloads is reduced.
In this paper, we describe parallels between this situation and dynamic power management as well as between the policy implemented by these CPUs and fixed-timeout device shutdown policies.
We show that the policy implemented by Intel CPUs is not optimal and describe potential better policies.
In particular, we present a mechanism to classify applications based on their likeliness to cause frequency reduction.
Our approach takes either the resulting classification information or information provided by the application and generates hints for the DVFS policy.
We show that faster frequency changes based on these hints are able to improve performance for a web server using the OpenSSL library.
\end{abstract}

\section{Introduction}
\label{sec:intro}

In recent years, performance became increasingly limited by power consumption as Dennard scaling has come to an end~\cite{taylor2012dark}.
The effect where the available power budget allows for different maximum frequencies depending on the number of cores is called dim silicon~\cite{huang2011scaling}.
The same effect also applies to different instruction mixes.
As different operations cause different switching activity on the chip, they consume different amounts of energy, so complex instructions have to be executed at a lower frequency.
Similarly, if unused parts of the chip are power-gated because they are not required by simpler operations, the resulting power savings can be used to increase the frequency.
The power budget is not only limited due to thermal constraints but also due to power supply limitations\footnote{In our tests, recent Intel CPUs have reported maximum current as the most common reason for frequency changes in AVX-512-heavy workloads.}, where even short-term transgressions could cause instability due to voltage drops.

As the large size of the SIMD registers used by recent SIMD instruction set extensions causes high power variation, recent CPUs have started to vary their frequency based on the workload to maximize performance under power budget constraints.
For example, Intel CPUs reduce their clock speed as soon as code containing AVX2 and AVX-512 instructions is executed~\cite{xeonscalableerrata}.
However, every frequency change causes some overhead~\cite{park2013accurate}, because the system has to wait for voltages to change\footnote{The frequency can only be increased when sufficient voltage is available, leading to frequency change delays and a resulting \enquote{underclocking loss}~\cite{park2013accurate}.} and clock signals to stabilize.
Therefore, even if no AVX2 and AVX-512 instructions are executed anymore, these CPUs delay increasing the clock speed\cite{optimizationmanual}.
This mechanism ensures that if the code continues executing these vectorized instructions shortly after, no excessive numbers of frequency changes are performed.

For some workloads, the delay causes overhead, though, as parts of the software which could be executed at higher frequency are needlessly slowed down.
For example, a simple benchmark using the nginx web server is slowed down by 10\% if the SSL library used by the web server is compiled with support for AVX-512, as the CPU frequency is reduced during AVX-512-heavy encryption and decryption, but the frequency change also affects the non-vectorized parts of the web server~\cite{krasnovdangers}.

A policy similar to this constant-delay policy is employed in the area of dynamic power management.
In this area, a similar trade-off is found, as disabling devices saves energy but incurs overhead both during shutdown and reactivation.
The widely-used \emph{fixed timeout} policy shuts down devices after a fixed delay~\cite{benini2000survey}, where the delay is usually equal to the \emph{break-even time} in order to improve worst-case power consumption~\cite{karlin1994competitive}.
In the area of power management, research has brought up a plethora of other shutdown strategies promising higher energy savings~\cite{benini2000survey} and has shown that input from the application can be used to further improve power efficiency~\cite{venkatachalam2005power}.
It is likely that similar approaches can be used to reduce DVFS overhead for partially power-intensive workloads.
In this work, we show that, in particular, input from the application can be used to predict whether immediate reclocking makes sense.
Our contributions are as follows:

\begin{table*}[t]
	\begin{center}
		\begin{tabular}{|r|c|c|c|c|c|}
			\hline
			Active cores & 1-2 cores & 3-4 cores & 5-8 cores & 9-12 cores & 13-16 cores \\
			\hline
			\hline
			Normal & \SI{3.7}{\giga\hertz} & \SI{3.5}{\giga\hertz} & \SI{3.4}{\giga\hertz} & \SI{3.1}{\giga\hertz} & \SI{2.8}{\giga\hertz} \\
			\hline
			AVX2 & \SI{3.6}{\giga\hertz} & \SI{3.4}{\giga\hertz} & \SI{3.1}{\giga\hertz} & \SI{2.6}{\giga\hertz} & \SI{2.4}{\giga\hertz} \\
			\hline
			AVX-512 & \SI{3.5}{\giga\hertz} & \SI{3.1}{\giga\hertz} & \SI{2.4}{\giga\hertz} & \SI{2.1}{\giga\hertz} & \SI{1.9}{\giga\hertz} \\
			\hline
		\end{tabular}
	\end{center}
	\caption{
		Maximum turbo frequency of the Intel Xeon Gold 6130 processor~\cite{wikichip6130}.
		The frequency reduction caused by AVX2 and AVX-512 instructions increases when more cores are active.
	}
	\label{tab:freqlevels}
\end{table*}

\begin{figure*}[t]
	\begin{center}
	\begin{tikzpicture}
		\begin{axis}[
			ybar,
			/pgf/bar width=5pt,
			ylabel style={align=center},
			ylabel={CPU time\\{}(normalized)},
			symbolic x coords={
				nginx+openssl,
				nginx+x265,
				blackscholes,
				fluidanimate,
				swaptions,
				vips,
				x264,
				apache,
				build-linux-kernel,
				git,
				mysqlslap,
				redis,
				sqlite
			},
			enlarge x limits=0.1,
			xtick=data,
			x tick label style={rotate=45,anchor=east},
			nodes near coords,
			every node near coord/.append style={color=black,rotate=90, anchor=east},
			ymin=0,
			ymax=1.6,
			width=14cm,
			height=4.3cm,
			point meta=explicit symbolic,
			xtick align=inside,
			major x tick style = {opacity=0},
			minor x tick num = 1,
			legend style={legend columns=3,at={(axis description cs:1.0,-0.6)}, anchor=east},
			axis background/.style={%
				preaction={
					path picture={
						\coordinate (parsectopleft) at ($ (axis cs:nginx+x265,1.9)!0.5!(axis cs:blackscholes,1.9) $);
						\coordinate (parsecbotleft) at ($ (axis cs:nginx+x265,0)!0.5!(axis cs:blackscholes,0) $);
						\draw [gray,sharp plot,dashed] (parsectopleft) -- (parsecbotleft);
						\coordinate (parsectopright) at ($ (axis cs:x264,1.9)!0.5!(axis cs:apache,1.9) $);
						\coordinate (parsecbotright) at ($ (axis cs:x264,0)!0.5!(axis cs:apache,0) $);
						\draw [gray,sharp plot,dashed] (parsectopright) -- (parsecbotright);
						\node (b) at (axis cs:swaptions,1.4) {Parsec};
						\node (b) at ($ (axis cs:git,1.4)!0.5!(axis cs:mysqlslap,1.4) $) {Phoronix Test Suite};
			}}},
		]
		\addplot[color=kitblue, fill=kitblue!75, error bars/.cd, y dir=both, y explicit] coordinates {
			(blackscholes,1)
			(fluidanimate,1)
			(nginx+x265,1)
			(nginx+openssl,1)
			(apache,1)
			(build-linux-kernel,1)
			(git,1)
			(mysqlslap,1)
			(redis,1)
			(sqlite,1)
			(swaptions,1)
			(vips,1)
			(x264,1)
		};
		\addplot[color=kitgreen, fill=kitgreen!75, error bars/.cd, y dir=both, y explicit] coordinates {
			(blackscholes,1.0305009670609735)
			(fluidanimate,1.0479568701236082)
			(nginx+x265,1.0661325443998315)
			(nginx+openssl,0.987599135455974)
			(apache,1.1071207777206167)
			(build-linux-kernel,1.0245829060485647)
			(git,0.9971475271064312)
			(mysqlslap,1.0563514631166315)
			(redis,1.0146497484963843)
			(sqlite,1.0637061689169138)
			(swaptions,1.030174827342295)
			(vips,1.0325687650950488)
			(x264,1.0417765611181158)
		};
		\addplot[color=kitbrown, fill=kitbrown!75, error bars/.cd, y dir=both, y explicit] coordinates {
			(blackscholes,1.0993230006326142)
			(fluidanimate,1.1111265721357355)
			(nginx+x265,1.2184270220754838)
			(nginx+openssl,1.1067079781336306)
			(apache,1.1891641518383906)
			(build-linux-kernel,1.1201097426066982)
			(git,1.0369975152488538)
			(mysqlslap,1.1300281054665602)
			(redis,1.114664749122852)
			(sqlite,1.152169203928504)
			(swaptions,1.0700976852405202)
			(vips,1.0749237263715143)
			(x264,1.1461191278167302)
		};
		\legend{AVX, AVX2, AVX-512}
		\end{axis}
	\end{tikzpicture}
	\end{center}
	\caption{
		CPU time required to run various benchmarks under the influence of different instruction sets to measure the impact of AVX frequency reduction.
	}
	\label{fig:overhead-with-ht}
\end{figure*}
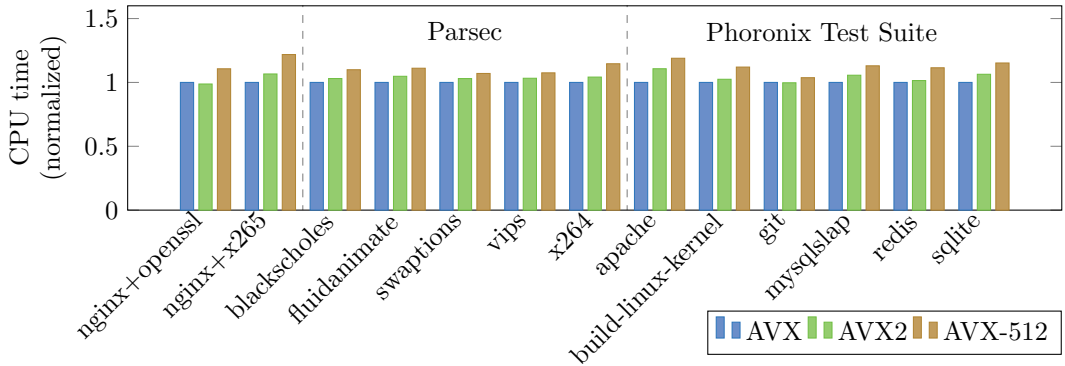
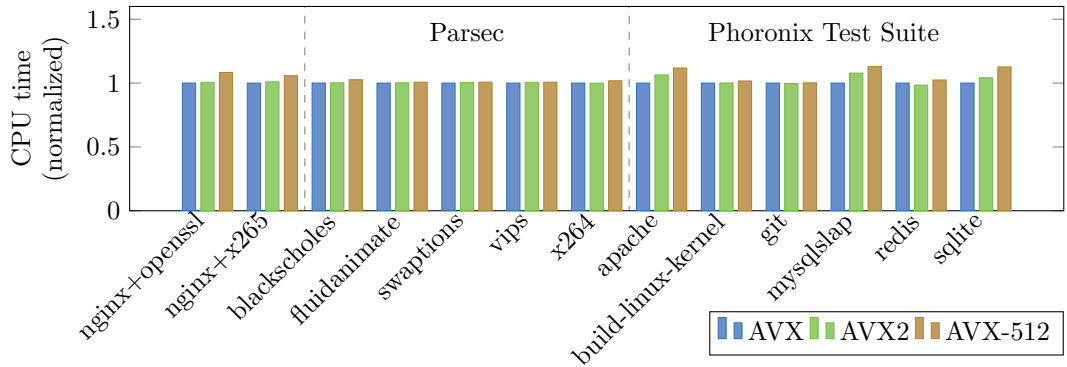
\begin{figure*}[t]
	\begin{center}
	\begin{tikzpicture}
		\begin{axis}[
			ybar,
			/pgf/bar width=5pt,
			ylabel style={align=center},
			ylabel={CPU time\\{}(normalized)},
			symbolic x coords={
				nginx+openssl,
				nginx+x265,
				blackscholes,
				fluidanimate,
				swaptions,
				vips,
				x264,
				apache,
				build-linux-kernel,
				git,
				mysqlslap,
				redis,
				sqlite
			},
			enlarge x limits=0.1,
			xtick=data,
			x tick label style={rotate=45,anchor=east},
			nodes near coords,
			every node near coord/.append style={color=black,rotate=90, anchor=east},
			ymin=0,
			ymax=1.6,
			width=14cm,
			height=4.3cm,
			point meta=explicit symbolic,
			xtick align=inside,
			major x tick style = {opacity=0},
			minor x tick num = 1,
			legend style={legend columns=3,at={(axis description cs:1.0,-0.6)}, anchor=east},
			axis background/.style={%
				preaction={
					path picture={
						\coordinate (parsectopleft) at ($ (axis cs:nginx+x265,1.9)!0.5!(axis cs:blackscholes,1.9) $);
						\coordinate (parsecbotleft) at ($ (axis cs:nginx+x265,0)!0.5!(axis cs:blackscholes,0) $);
						\draw [gray,sharp plot,dashed] (parsectopleft) -- (parsecbotleft);
						\coordinate (parsectopright) at ($ (axis cs:x264,1.9)!0.5!(axis cs:apache,1.9) $);
						\coordinate (parsecbotright) at ($ (axis cs:x264,0)!0.5!(axis cs:apache,0) $);
						\draw [gray,sharp plot,dashed] (parsectopright) -- (parsecbotright);
						\node (b) at (axis cs:swaptions,1.4) {Parsec};
						\node (b) at ($ (axis cs:git,1.4)!0.5!(axis cs:mysqlslap,1.4) $) {Phoronix Test Suite};
			}}},
		]
		\addplot[color=kitblue, fill=kitblue!75, error bars/.cd, y dir=both, y explicit] coordinates {
			(blackscholes,1)
			(fluidanimate,1)
			(nginx+x265,1)
			(nginx+openssl,1)
			(apache,1)
			(build-linux-kernel,1)
			(git,1)
			(mysqlslap,1)
			(redis,1)
			(sqlite,1)
			(swaptions,1)
			(vips,1)
			(x264,1)
		};
		\addplot[color=kitgreen, fill=kitgreen!75, error bars/.cd, y dir=both, y explicit] coordinates {
			(blackscholes,1.0016854871785301)
			(fluidanimate,1.0021996252338796)
			(nginx+x265,1.0099157911329861)
			(nginx+openssl,1.005252466053777)
			(apache,1.063889597523588)
			(build-linux-kernel,1.0004597591151683)
			(git,0.9958381496715788)
			(mysqlslap,1.0779309600134774)
			(redis,0.9838276057321378)
			(sqlite,1.0412665085621042)
			(swaptions,1.003225747625358)
			(vips,1.0032861566785727)
			(x264,0.9992729799611385)
		};
		\addplot[color=kitbrown, fill=kitbrown!75, error bars/.cd, y dir=both, y explicit] coordinates {
			(blackscholes,1.0262038799716826)
			(fluidanimate,1.0062495284485728)
			(nginx+x265,1.057762817274907)
			(nginx+openssl,1.0826257572002929)
			(apache,1.1177798639449468)
			(build-linux-kernel,1.0162686828010652)
			(git,1.0016761796136444)
			(mysqlslap,1.1289166565703017)
			(redis,1.0240709547970706)
			(sqlite,1.1261298996824225)
			(swaptions,1.0075415655909874)
			(vips,1.00665867494822)
			(x264,1.0176460019252347)
		};
		\legend{AVX, AVX2, AVX-512}
		\end{axis}
	\end{tikzpicture}
	\end{center}
	\caption{
		CPU time required for the experiment shown in Figure~\ref{fig:overhead-with-ht} on a system with hyperthreading disabled.
		Note that benchmarks with few context switches do not suffer from frequency changes anymore once hyperthreading is disabled, whereas benchmarks with frequent switches between AVX and non-AVX code (i.e., heterogeneous programs with short AVX-heavy phases as well as workloads consisting of an interactive service and an AVX-heavy background task) still suffer from the frequency change delay.
		Note that such workloads are often latency-critical and therefore particularly suffer from degraded performance.
	}
	\label{fig:overhead-without-ht}
\end{figure*}

\begin{itemize}
	\item We describe the parallels between DVFS in dim silicon scenarios and dynamic power management.
		The duality allows to apply research from the area of dynamic power management to the former.
	\item We determine the frequency change cost on a current server system and calculate the break-even time for frequency changes.
		We use this result to show how the delay specified by Intel does not provide optimal worst-case behavior.
	\item We show that application knowledge about execution phases or the instruction types used by individual processes can be used to improve performance by passing hints about future instruction set usage to the DVFS policy.
		We validate this finding through simulation of different DVFS policies on a web server workload.
	\item We describe a mechanism to determine at runtime whether individual processes will trigger frequency reductions due to their usage of power-intensive instructions.
		Unlike existing approaches, our design can reliably distinguish between all three frequency levels provided by current Intel CPUs.
		This information can be used as input for an improved DVFS policy to trigger frequency changes during context switches.
\end{itemize}

\section{Effects of AVX2 and AVX-512}
\label{sec:avxeffects}

Starting with the Haswell microarchitecture which introduced the AVX2 instruction set, Intel introduced a separate maximum frequency for AVX2-intensive code segments~\cite{hackenberg2015energy}.
The Skylake microarchitecture added AVX-512 instructions and a third AVX-512 frequency level~\cite{schone2019energy}.
Table~\ref{tab:freqlevels} shows the maximum turbo frequency for the Intel Xeon Gold 6130 server processor.
The maximum frequency depends both on the number of active cores -- with larger numbers of active cores requiring larger frequency reduction -- as well as on the type of instructions executed.
AVX-512 causes a particularly large frequency reduction due to the complexity of operations on 512-bit vectors.
As described above, the reduced frequency is maintained longer than necessary to prevent excessive reclocking overhead.

There are two situations where this delay can cause the frequency reduction to negatively affect unrelated non-AVX code and cause a significant performance reduction.
First, on a system with simultaneous multithreading (SMT), if one of the hardware threads causes the frequency of the physical core to be reduced, the other hardware threads on the same core also execute at lower frequency even if their code is not as energy-intensive~\cite{li2019corescheduling}.
Second, in heterogeneous applications consisting of power-intensive and less power-intensive parts -- or if the OS frequently switches between power-intensive and less power-intensive tasks -- the delay before increasing the frequency causes reduced performance for the less power-intensive code~\cite{gottschlag19sfma}.

As an example for the latter, previous work describes overhead caused by AVX-512{} in a web server workload, where the nginx web server provides up to 10\% lower performance when the SSL library uses cryptography primitives implemented with AVX-512 instructions, because unrelated web server code is slowed down following calls into the SSL library~\cite{krasnovdangers}.
We replicated this experiment, the result is shown in Figure~\ref{fig:overhead-with-ht} alongside other experiments with workloads consisting of multiple different processes to show that the performance impact is also present in such scenarios.
For these other experiments, we execute different non-AVX workloads while concurrently executing the x265 video encoder configured to use AVX, AVX2, or AVX-512 instructions.
The experiments are conducted on a system with an Intel Xeon Gold 6130 processor.

Our first multi-process experiment determines the impact on an interactive web server workload:
We executed the nginx web server alongside the x265 video encoder and configured the wrk2 client to generate a fixed number of requests to the web server.
This setup imitates the scenario where a web server is not fully utilized and the remaining CPU time is used for background batch tasks.
Figure~\ref{fig:overhead-with-ht} shows the normalized CPU time required by the nginx web server to serve a unencrypted static file (\enquote{nginx+x265}).
The results show a 6.6\% performance impact when the background process uses AVX2 instructions and a 21.8\% performance impact for AVX-512.
As the web server is not operating at 100\% utilization, the background process is often executed inbetween two consecutive requests or is executed in parallel on the other hardware thread of the same core, causing a particularly large performance impact.

To show that the problem affects both interactive and batch workloads, we also execute various benchmarks from the Parsec~\cite{bienia11benchmarking} benchmark suite and the Phoronix Test Suite (PTS)~\cite{phoronixtestsuite} benchmarks in parallel to the x265 video encoder.
As shown in Figure~\ref{fig:overhead-with-ht}, all these benchmarks are also affected by the frequency changes caused by x265.
The Parsec benchmarks experience an average performance reduction by 10.0\% for AVX-512.
Similarly, the PTS benchmarks are slowed down by 12.4\%.


As described above, one major mechanism for slowdown that is targeted by other approaches~\cite{li2019corescheduling} is that software on one hardware thread slows down other hardware threads of the same core.
To show that some of the slowdown is also experienced on systems without hyperthreading, we repeat all the benchmarks on a system with hyperthreading disabled.
The results of this experiment are shown in Figure~\ref{fig:overhead-without-ht} and show that CPU-intensive non-interactive workloads are not significantly slowed down anymore once hyperthreading is disabled as the system does not switch between the processes often enough for frequency change delays to have a significant effect.
For example, on a system with the default Linux CFS scheduler, we observe only one context switch every 10 to \SI{20}{\milli\second} for the blackscholes workload whereas frequency increases are only delayed by less than one millisecond.
Although disabling hyperthreading reduces the performance of the system and is therefore not a viable technique against the overhead caused by AVX-heavy code in these scenarios, other techniques such as core specialization~\cite{gottschlag19sfma} and core scheduling~\cite{li2019corescheduling} can make sure that whenever possible either both hyperthreads are executing AVX-intensive code or none of them is.

Overhead caused by hyperthreading is out of the scope of this paper, though.
Instead, the goal of our approach is to reduce the overhead in applications which periodically execute short sections of AVX-512 or AVX2 code as well as in workloads which frequently switch between AVX-512 or AVX2 and non-AVX applications on a single core.
From the benchmarks shown in Figure~\ref{fig:overhead-without-ht}, an example for the former is the nginx/OpenSSL benchmark, which executes AVX-512 instructions only when OpenSSL functions are called.
The nginx/x265 benchmark as well as the Apache, MySQL and SQLite benchmarks from PTS, instead, trigger frequent context switches between the AVX-512-enabled background task and the benchmarked application and are therefore examples for the latter behaviour.
These types of benchmarks are the benchmarks which show overhead even when hyperthreading is disabled:
For AVX-512, the nginx benchmarks are slowed down by 7.0\% on average, whereas the three PTS benchmarks are slowed down by 12.4\% on average.

Due to the frequent switches between AVX-512/AVX2 and non-AVX code during these workloads, the upclocking delay implemented by the CPU's existing hardware DVFS policy is the main source for the overhead caused by AVX instructions.
To isolate this overhead source and to demonstrate that improved DVFS policies are able to mitigate its effects, we conduct all further experiments in this paper with hyperthreading disabled.
The assumption of CPUs without hyperthreading significantly simplifies the design of some parts of our approach.
This does not mean that improved DVFS policies are inherently ineffective on systems with hyperthreading, although more research has to be conducted to identify appropriate heuristics for improved DVFS decisions.

\section{Parallels to Dynamic Power Management}
\label{sec:analysis}

As described above, the complex frequency behavior of modern CPUs stems from the fact that it is not economically viable to cool modern CPUs when they are executing power-intensive code at their maximum frequency~\cite{huang2011scaling}.
Instead, available thermal headroom is used to temporarily use higher frequencies (a form of computational sprinting~\cite{raghavan2012computational}).
In this scenario, the more the energy consumption per instruction varies, the higher is the thermal headroom for code executing simple instructions.
Therefore, modern Intel CPUs use different turbo frequencies for different types of code, with AVX2 and AVX-512 instructions triggering a transition to significantly lower frequency levels~\cite{schone2019energy}.
As shown by the registers provided by these CPUs to determine the reason for frequency changes, not only thermal headroom plays a factor for these frequency reductions, though:
The power dissipation of the chip correlates with the current required from the power supply, and frequency changes are also required to prevent voltage drops due to increased current draw.

The frequency changes required to use the available headroom come at a cost.
For example, Mazouz et al. have measured the cost of a single frequency change to be approximately \SI{10}{\micro\second} on an Intel Ivy Bridge system~\cite{mazouz2014evaluation} and our own experiments presented in Section~\ref{sec:freqchangecost} arrive at a similar cost (between \SI{9}{\micro\second} and \SI{19}{\micro\second}) on more recent Skylake server CPUs.
Therefore, increasing the frequency to use thermal headroom is only viable if the higher frequency can be applied long enough that the performance improvement makes up for the frequency change overhead.
This trade-off is similar to the problem of \emph{dynamic power management} where devices are temporarily switched off or transitioned to a low-power state in order to save energy~\cite{benini2000survey}.
Here, the energy cost for the state transition means that switching devices off for only short periods of time is frequently unviable.
As the operating system, however, does not know how long a device is going to stay unused, it is in general not possible to determine in advance whether shutting a device off is going to result in a net improvement.

In the area of dynamic power management, significant effort has gone into developing heuristic approaches to guess when to shutdown devices~\cite{benini2000survey}.
One metric to measure the quality of heuristic approaches is their \emph{competitiveness} in a worst-case scenario.
The competitiveness is the worst-case ratio between the energy required by the approach compared to the energy required by an oracle policy that can determine in advance whether shutting off a device is viable.
Karlin et al.~\cite{karlin1994competitive} showed at most 2-competitiveness (meaning that the approach uses at most twice as much energy) is possible for deterministic algorithms.
In dynamic power management, 2-competitiveness can be achieved by switching a device off after a fixed timeout.
When that timeout equals the \emph{break-even time} (i.e., the time of inactivity during with the low-power state would have made up for the transition costs), the device uses at most twice as much energy if it wakes up directly after being sent to a low-power state.
Intel CPUs show a very similar behavior as they delay increasing the frequency by a fixed timeout after the CPU has stopped executing any AVX instructions~\cite{schone2019energy}.
However, the fixed delay is not optimal in terms of competitiveness because, as we show in Section~\ref{sec:competitiveness}, DVFS has wildly varying break-even times in different scenarios.
Neither is the DVFS policy implemented by current Intel CPUs optimal for real-world workloads as we show in Section~\ref{sec:dvfs-policy-eval}.

There are approaches that can, depending on the situation, perform better than simple heuristic approaches.
For example, applications can give hints about expected future behavior to let the OS perform better informed decisions~\cite{lu2002power} or the OS can use the deadlines of I/O requests to change the device usage pattern to save more energy~\cite{weissel2002cooperative}
Both these approaches can be applied to DVFS policies in dim silicon scenarios.
In this paper, we show an example for the former approach.
As software developers often know whether the application is going to execute no power-intensive code -- i.e., no AVX2 and AVX-512 -- in the near future, that information can be used by the CPU to forego the frequency change delay and immediately change frequencies for improved performance.

\section{Behavior of Intel CPUs}
\label{sec:competitiveness}

According to the optimization manual, recent Intel CPUs implement a fixed-timeout policy where the CPU waits approximately \SI{2}{\milli\second} after the last section of AVX-intensive code before increasing the frequency again~\cite[p. 2-13]{optimizationmanual}.
In addition, before lowering the frequency, the core requests a power license from the package control unit (PCU) which takes up to \SI{500}{\micro\second} before granting the license.
However, as shown by Schöne et al.~\cite{schone2019energy}, the behavior of the hardware does not match the documentation.
Instead, the processor waits for a significantly shorter timeout (approx. \SI{670}{\micro\second} as measured in our experiments) before upclocking.
We were able to confirm the observed behavior on a system with an Intel Core i9-7940X, where we measured the delay for frequency changes when executing sections of code consisting of scalar, AVX2, or AVX512 instructions.
Note that frequency reduction is triggered almost immediately when AVX2 or AVX-512 instructions are executed, as required to prevent excessive power consumption.

The upclocking delay is constant independent from the number of cores in use.
As described in the last section, maximum competitiveness in worst-case scenarios is reached when the timeout equals the break-even time, but the break-even time depends not only on the cost for the frequency transition but also on the performance advantage at a higher frequency.
In this case, the frequency change is higher if more cores are active~\cite{xeonscalableerrata}, so the performance overhead for downclocking is higher and the break-even time is shorter when more cores are active.
Therefore, the policy implemented by Intel does not provide maximum competitiveness.
To show the potential for improved timeout-based policies, the following sections describe experiments to determine both frequency transition overhead as well as performance impact for different situations to determine the corresponding break-even times.

\subsection{Cost of Frequency Changes}
\label{sec:freqchangecost}

\begin{figure}[t]
	\begin{center}
	\begin{tikzpicture}
		\begin{axis}[
			xlabel=Active cores,
			ylabel=Overhead ($\mu$s),
			height=5cm,
			width=8cm,
			ymax=55,
			ymin=0
		]
		\addplot[color=kitblue,mark=x, error bars/.cd, y dir=both, y explicit] coordinates {
			(1, 18.3831) -= (0, 2.11772) += (0, 2.11772)
			(2, 18.3701) -= (0, 5.75913) += (0, 5.75913)
			(3, 18.7486) -= (0, 3.57153) += (0, 3.57153)
			(4, 18.9329) -= (0, 9.45622) += (0, 9.45622)
			(5, 17.7967) -= (0, 2.39951) += (0, 2.39951)
			(6, 17.7583) -= (0, 2.23907) += (0, 2.23907)
			(7, 17.1102) -= (0, 6.22886) += (0, 6.22886)
			(8, 17.8264) -= (0, 21.4344) += (0, 21.4344)
			(9, 15.7571) -= (0, 1.85543) += (0, 1.85543)
			(10, 15.7666) -= (0, 2.52449) += (0, 2.52449)
			(11, 15.7953) -= (0, 2.44806) += (0, 2.44806)
			(12, 15.8192) -= (0, 7.59365) += (0, 7.59365)
			(13, 16.1483) -= (0, 2.55338) += (0, 2.55338)
			(14, 16.3903) -= (0, 1.80104) += (0, 1.80104)
			(15, 16.565) -= (0, 1.96258) += (0, 1.96258)
			(16, 16.3513) -= (0, 2.47415) += (0, 2.47415)
		};
		\addplot[color=kitgreen,mark=o, error bars/.cd, y dir=both, y explicit] coordinates {
			(1, 27.8095) -= (0, 2.45606) += (0, 2.45606)
			(2, 28.0433) -= (0, 9.54818) += (0, 9.54818)
			(3, 26.9738) -= (0, 8.88508) += (0, 8.88508)
			(4, 26.7712) -= (0, 25.024) += (0, 25.024)
			(5, 22.7115) -= (0, 1.76558) += (0, 1.76558)
			(6, 22.9779) -= (0, 3.05395) += (0, 3.05395)
			(7, 22.8565) -= (0, 2.3046) += (0, 2.3046)
			(8, 24.2017) -= (0, 20.2542) += (0, 20.2542)
			(9, 23.28) -= (0, 1.90465) += (0, 1.90465)
			(10, 23.299) -= (0, 2.15714) += (0, 2.15714)
			(11, 23.3228) -= (0, 4.93699) += (0, 4.93699)
			(12, 24.0869) -= (0, 11.1654) += (0, 11.1654)
			(13, 22.9453) -= (0, 1.92664) += (0, 1.92664)
			(14, 22.2663) -= (0, 3.84499) += (0, 3.84499)
			(15, 23.1522) -= (0, 1.99213) += (0, 1.99213)
			(16, 23.1068) -= (0, 1.95261) += (0, 1.95261)
		};
		\addplot[color=kitbrown,mark=*, error bars/.cd, y dir=both, y explicit] coordinates {
			(1, 17.0681) -= (0, 5.72346) += (0, 5.72346)
			(2, 17.9898) -= (0, 11.3278) += (0, 11.3278)
			(3, 16.609) -= (0, 7.57117) += (0, 7.57117)
			(4, 17.199) -= (0, 23.2871) += (0, 23.2871)
			(5, 12.8568) -= (0, 3.81768) += (0, 3.81768)
			(6, 12.8128) -= (0, 4.22659) += (0, 4.22659)
			(7, 12.383) -= (0, 5.63913) += (0, 5.63913)
			(8, 12.6608) -= (0, 18.1051) += (0, 18.1051)
			(9, 14.4816) -= (0, 5.84171) += (0, 5.84171)
			(10, 14.4785) -= (0, 3.57992) += (0, 3.57992)
			(11, 13.7233) -= (0, 4.89692) += (0, 4.89692)
			(12, 13.7987) -= (0, 14.105) += (0, 14.105)
			(13, 13.2402) -= (0, 4.35107) += (0, 4.35107)
			(14, 12.97) -= (0, 4.79027) += (0, 4.79027)
			(15, 13.1401) -= (0, 5.01173) += (0, 5.01173)
			(16, 14.3547) -= (0, 3.29653) += (0, 3.29653)
		};
		\legend{Scalar $\rightarrow$ AVX2, Scalar $\rightarrow$ AVX-512, AVX2 $\rightarrow$ AVX-512}
		\end{axis}
	\end{tikzpicture}
	\end{center}
	\caption{
		Overhead when the frequency is reduced, measured as the mean of 1000 runs.
		The error bars indicate the standard deviation.
		The overhead seems to vary slightly based on the number of active cores and on the resulting frequencies.
		Note that a transition from scalar to AVX-512 frequencies incurs two separate frequency transitions.
	}
	\label{fig:freq-reduction-overhead-results}
\end{figure}
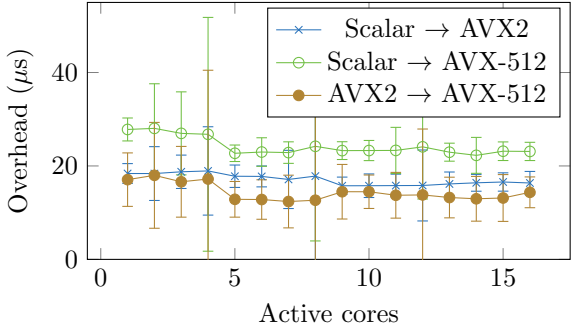
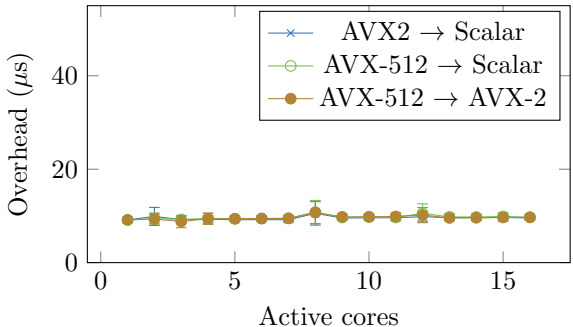
\begin{figure}[t]
	\begin{center}
	\begin{tikzpicture}
		\begin{axis}[
			xlabel=Active cores,
			ylabel=Overhead ($\mu$s),
			height=5cm,
			width=8cm,
			ymax=55,
			ymin=0
		]
		\addplot[color=kitblue,mark=x, error bars/.cd, y dir=both, y explicit] coordinates {
			(1, 9.12968) -= (0, 0.371133) += (0, 0.371133)
			(2, 9.89633) -= (0, 1.92646) += (0, 1.92646)
			(3, 9.19064) -= (0, 0.564885) += (0, 0.564885)
			(4, 9.27862) -= (0, 0.789639) += (0, 0.789639)
			(5, 9.16189) -= (0, 0.329754) += (0, 0.329754)
			(6, 9.20446) -= (0, 0.353613) += (0, 0.353613)
			(7, 9.16998) -= (0, 0.500273) += (0, 0.500273)
			(8, 10.5573) -= (0, 2.5259) += (0, 2.5259)
			(9, 9.54123) -= (0, 0.396032) += (0, 0.396032)
			(10, 9.59297) -= (0, 0.389191) += (0, 0.389191)
			(11, 9.61289) -= (0, 0.667445) += (0, 0.667445)
			(12, 9.84082) -= (0, 0.943753) += (0, 0.943753)
			(13, 9.61627) -= (0, 0.341981) += (0, 0.341981)
			(14, 9.63797) -= (0, 0.415312) += (0, 0.415312)
			(15, 9.66503) -= (0, 0.42425) += (0, 0.42425)
			(16, 9.56553) -= (0, 0.381955) += (0, 0.381955)
		};
		\addplot[color=kitgreen,mark=o, error bars/.cd, y dir=both, y explicit] coordinates {
			(1, 9.15986) -= (0, 0.499383) += (0, 0.499383)
			(2, 9.43172) -= (0, 1.17339) += (0, 1.17339)
			(3, 9.24175) -= (0, 0.811743) += (0, 0.811743)
			(4, 9.42288) -= (0, 1.11563) += (0, 1.11563)
			(5, 9.36885) -= (0, 0.619109) += (0, 0.619109)
			(6, 9.43908) -= (0, 0.595626) += (0, 0.595626)
			(7, 9.49606) -= (0, 0.643047) += (0, 0.643047)
			(8, 10.815) -= (0, 2.49958) += (0, 2.49958)
			(9, 9.66177) -= (0, 0.627242) += (0, 0.627242)
			(10, 9.7525) -= (0, 0.691539) += (0, 0.691539)
			(11, 9.66297) -= (0, 0.629896) += (0, 0.629896)
			(12, 10.5623) -= (0, 2.02091) += (0, 2.02091)
			(13, 9.78831) -= (0, 0.647968) += (0, 0.647968)
			(14, 9.76628) -= (0, 0.672887) += (0, 0.672887)
			(15, 9.88321) -= (0, 0.815622) += (0, 0.815622)
			(16, 9.72921) -= (0, 0.648825) += (0, 0.648825)
		};
		\addplot[color=kitbrown,mark=*, error bars/.cd, y dir=both, y explicit] coordinates {
			(1, 9.13772) -= (0, 0.596498) += (0, 0.596498)
			(2, 9.32953) -= (0, 1.00844) += (0, 1.00844)
			(3, 8.79505) -= (0, 1.31756) += (0, 1.31756)
			(4, 9.40442) -= (0, 1.22505) += (0, 1.22505)
			(5, 9.40862) -= (0, 0.543706) += (0, 0.543706)
			(6, 9.44813) -= (0, 0.545447) += (0, 0.545447)
			(7, 9.46255) -= (0, 0.523241) += (0, 0.523241)
			(8, 10.6957) -= (0, 2.35805) += (0, 2.35805)
			(9, 9.80298) -= (0, 0.58775) += (0, 0.58775)
			(10, 9.81613) -= (0, 0.497965) += (0, 0.497965)
			(11, 9.88233) -= (0, 0.812184) += (0, 0.812184)
			(12, 10.2016) -= (0, 1.58667) += (0, 1.58667)
			(13, 9.55865) -= (0, 0.633994) += (0, 0.633994)
			(14, 9.59252) -= (0, 0.662044) += (0, 0.662044)
			(15, 9.65426) -= (0, 0.809789) += (0, 0.809789)
			(16, 9.68098) -= (0, 0.55307) += (0, 0.55307)
		};
		\legend{AVX2 $\rightarrow$ Scalar, AVX-512 $\rightarrow$ Scalar, AVX-512 $\rightarrow$ AVX-2}
		\end{axis}
	\end{tikzpicture}
	\end{center}
	\caption{
		Overhead when the frequency is increased.
		No variation based on the number of active cores can be observed.
	}
	\label{fig:freq-increase-overhead-results}
\end{figure}

One factor required to determine the break-even time is the frequency change overhead:
If the cost of individual frequency changes increases, more time between consecutive changes is required in order to make up for the overhead.
For the Intel Ivy Bridge architecture, Mazouz et al. determined that a CPU is stopped for approximately \SI{10}{\micro\second} during a frequency change~\cite{mazouz2014evaluation}.
This pause is required to allow the new frequency to stabilize~\cite{park2013accurate}.
However, in particular in the case of frequency changes caused by AVX instructions, additional factors increase the overall overhead.
Therefore, and because our systems use a newer CPU architecture than the one considered by Mazouz et al., we measure the overhead of frequency changes on a system with an Intel Xeon Gold 6130 CPU.

To measure the overhead due to frequency reduction caused by AVX2 and AVX-512 instructions, we execute the same amount of such instructions twice, once when the system is already at the appropriate frequency, and once when it executes at a higher frequency and the code triggers a frequency change.
The overhead of the frequency change can be calculated as the difference of the two runtimes.
The results of this experiment for all combinations of scalar, AVX2, and AVX-512 instructions are shown in Figure~\ref{fig:freq-reduction-overhead-results}, which shows significantly higher overhead than measured by Mazouz et al.~\cite{mazouz2014evaluation}. 
For example, a transition from the maximum frequency to the AVX2 frequency level takes \SI{17}{\micro\second} on average, whereas a transition to the AVX-512 frequency level takes \SI{24}{\micro\second}.
The reason for this increased overhead is likely the reduced IPC due to additional throttling before the frequency switch is complete\cite{downs20gathering}.
As AVX2 and AVX-512 instructions would draw excessive power at the previous higher frequency, the system temporarily employs throttling to reduce power consumption~\cite{bonen2017performing}.
Note that the overhead appears to vary slightly for the different frequencies and frequency differences caused by different numbers of active cores.

Measuring the overhead of frequency increases is slightly more complex due to the large -- and, in our experiment, somewhat variable -- delay before the system restores the non-AVX frequency level.
In this case, we employ the technique employed by Mazouz et al. to determine frequency change costs~\cite{mazouz2014evaluation} as we start at a system running at either AVX2 or AVX-512 frequencies and repeatedly execute a short code section which consists of instructions allowing a higher frequency.
We measure the runtime of the code section each time, so that frequency changes are shown as spikes in the measured runtime.
As other sources such as the activation of additional cores can trigger additional reduction of the maximum frequency, we simply assume that the first frequency change is the one triggered by the lack of AVX2 and AVX-512 instructions and discard any further runtime spikes.
The size of the spike is assumed to be the overhead of the frequency change, which is plotted in Figure~\ref{fig:freq-increase-overhead-results}.
The results closely match those of Mazouz et al.~\cite{mazouz2014evaluation} and show no variation based on the absolute frequency of the core or the magnitude of the frequency change, both of which vary with the number of active cores.
Note, however, that this experiment does not consider the performance loss due to the system temporarily executing at a lower frequency while the voltage is ramped up to the level required for the frequency change~\cite{park2013accurate}.
For many dynamic power management approaches, state changes can be predicted in advance, so voltage changes can likely be conducted speculatively, removing the need for such additional delays.
For example, for fixed-timeout policies, the timeout can be slightly reduced accordingly.

\subsection{Performance Versus Frequency}

\pgfplotstableread{frequency-vs-performance.dat}{\loadeddata}

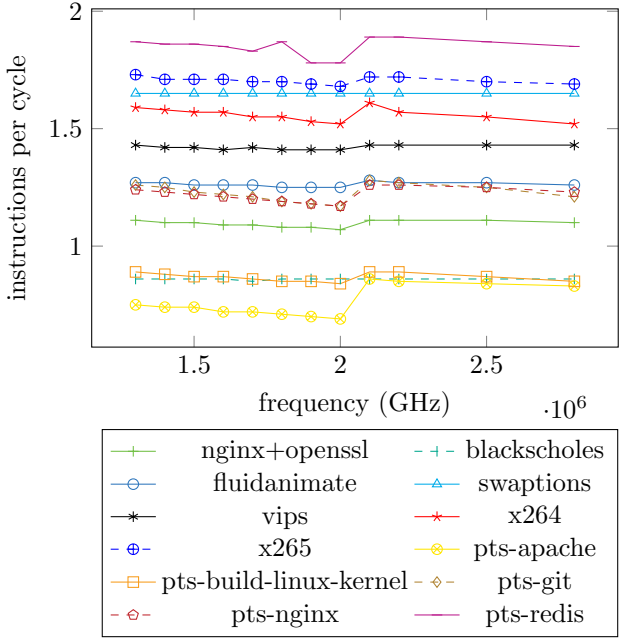
\begin{figure}[t]
	\begin{center}
	\begin{tikzpicture}
	\begin{axis}[
		scale only axis,
		xlabel={frequency (GHz)},
		ylabel={instructions per cycle},
		width=7cm,
		height=4.5cm,
		legend style={legend columns=2,at={(axis description cs:1.0,-0.55)}, anchor=east},
	]
	\definecolor{kitgreenexcl}{cmyk}{1.0,  0.0,  0.6, 0.0}
	\definecolor{kitblue}	 {cmyk}{0.8,  0.5,  0.0, 0.0}
	\definecolor{kitgreen}	{cmyk}{0.6,  0.0,  1.0, 0.0}
	\definecolor{kityellow}   {cmyk}{0.0,  0.05, 1.0, 0.0}
	\definecolor{kitorange}   {cmyk}{0.0,  0.45, 1.0, 0.0}
	\definecolor{kitbrown}	{cmyk}{0.35, 0.5,  1.0, 0.0}
	\definecolor{kitred}	  {cmyk}{0.25, 1.0,  1.0, 0.0}
	\definecolor{kitpurple}   {cmyk}{0.25, 1.0,  0.0, 0.0}
	\definecolor{kitcyan}	 {cmyk}{0.9,  0.05, 0.0, 0.0}
		\foreach \startindex/\endindex/\plotlabel/\plotmark/\plotcolor/\plotstyle in {
			24/ 35/ nginx+openssl/ +/ kitgreen/ solid,
			0/ 11/ blackscholes/ |/ kitgreenexcl/ dashed,
			12/ 23/ fluidanimate/ o/ kitblue/ solid,
			108/ 119/ swaptions/ triangle/ kitcyan/ solid,
			120/ 131/ vips/ asterisk/ black/ solid,
			132/ 143/ x264/ star/ red/ solid,
			144/ 155/ x265/ oplus/ blue/ dashed,
			36/ 47/ pts-apache/ otimes/ kityellow/ solid,
			48/ 59/ pts-build-linux-kernel/ square/ kitorange/ solid,
			60/ 71/ pts-git/ diamond/ kitbrown/ dashed,
			72/ 83/ pts-nginx/ pentagon/ kitred/ dashed,
			84/ 95/ pts-redis/ -/ kitpurple/ solid
		}{
			\edef\temp{\noexpand\addplot+[mark=\plotmark,mark options=solid,color=\plotcolor,\plotstyle,select coords between index={\startindex}{\endindex}] table[x=freq,y=ipc] {\noexpand\loadeddata};}
			\temp
			\addlegendentryexpanded{\plotlabel}
		}
	\end{axis}
	\end{tikzpicture}
	\end{center}
	\caption{
		IPC of various parsec and PTS benchmarks as well as the nginx/OpenSSL workload described in Section~\ref{sec:avxeffects} when executed at different frequencies.
		In the monotome region between \SI{2.1}{\giga\hertz} and \SI{2.8}{\giga\hertz} the benchmarks show little IPC variation.
		The step between \SI{2.0}{\giga\hertz} and \SI{2.1}{\giga\hertz} is likely because by either memory or bus frequency scaling in relation to the frequency of the cores.
	}
	\label{fig:frequency-vs-performance}
\end{figure}

The break-even time for frequency changes depends not only on the overhead for frequency transitions but also on the relative performance advantage due to the higher frequency.
Whereas the performance of CPU-bound tasks is nearly proportional to the CPU frequency, the same is not true for memory-heavy workloads as the memory latency is independent from the CPU frequency.
In this work, to simplify the prototype, we assume the former.

The result of this simplification is that the break-even time is underestimated for memory-heavy applications.
To quantify this error for the workloads used in this paper, we executed most of the individual applications described in Section~\ref{sec:avxeffects} -- nginx, x265, the Parsec benchmarks and the PTS benchmarks with the exception of mysql due to the long execution time of the corresponding benchmark and sqlite due to its particularly I/O-heavy nature -- at different frequencies and measured the instructions per cycle (IPC).
We executed the applications at frequencies between \SI{2.8}{\giga\hertz} and \SI{1.3}{\giga\hertz} on a system with a 16-core Intel Xeon Gold 6300 processor.
We configured the application to use all cores of the system except for the nginx server benchmark where we allocated three cores to the HTTP request generator\footnote{x265 failed to fully saturate all cores due to inter-thread dependencies.}.
Maximizing the number of active cores should maximize the working set of the application and should therefore maximize the impact of memory accesses on performance.

Figure~\ref{fig:frequency-vs-performance} shows the results of this experiment.
Counterintuitively, IPC consistently improves when the frequency is increased from \SI{2.0}{\giga\hertz} to \SI{2.1}{\giga\hertz} -- we assume this is due to the chip adapting either memory or bus frequency to the core frequency.
For all other frequency ranges, higher frequency correlates with lower IPC.
When comparing the IPC at \SI{2.1}{\giga\hertz} and \SI{2.8}{\giga\hertz}, the biggest difference was found for x264 which had 5.9\% higher IPC at \SI{2.1}{\giga\hertz}.
This IPC difference would translate into a error of 5.9\% during break-even time calculation, which is likely low enough for the simplified model to be viable for this workload.

The reason for the low IPC changes is found in the low cache miss rates for all these applications:
The workloads trigger at most 2.03 last-level cache misses per 1000 instructions (in the case of PTS build-linux-kernel).

Note that our simulation to show the viability of improved DVFS policies in Section~\ref{sec:dvfs-policy-eval} also uses the simplified performance model.
However, as our experiment shows, the resulting error is negligible and does not influence our conclusions.
The simulation uses the nginx web server with the configuration marked as \enquote{nginx+openssl} in Figure~\ref{fig:frequency-vs-performance}.
In this configuration, the nginx web server showed less than 1\% IPC difference between \SI{2.1}{\giga\hertz} and \SI{2.8}{\giga\hertz}.

Workloads with higher cache miss ratios than the benchmarks shown in Figure~\ref{fig:frequency-vs-performance} can show lower correlation between performance and frequency~\cite{hebbar2019impact}.
While we show that improved DVFS policies in general have the potential to improve performance for workloads involving AVX2 and AVX-512 code, our simplified linear model might not be sufficient for these workloads in practice.
Concrete DVFS policy implementations for such workloads might require a better prediction of the performance at different frequencies to make decisions on whether to change the CPU frequency or not.
Such predictions can be made, for example, by using performance counters to determine the impact of frequency changes on the number of stall cycles~\cite{keramidas2010interval}.
Further research has to be conducted to show whether DVFS policies based on such approaches are viable and provide a significant performance advantage for a wider range of workloads.

\subsection{Break-Even Time}
\label{sec:break-even-time}

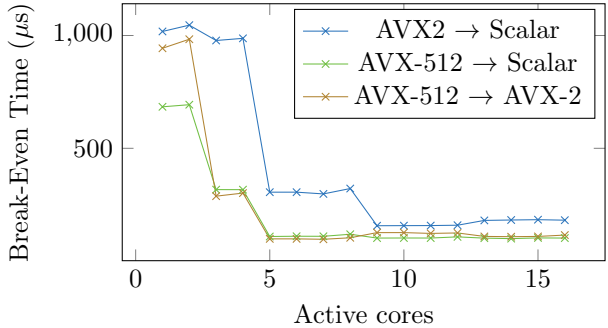
\begin{figure}[t]
	\begin{center}
	\begin{tikzpicture}
		\begin{axis}[
			xlabel=Active cores,
			ylabel=Break-Even Time ($\mu$s),
			height=5cm,
			width=8cm,
		]
		\addplot[color=kitblue,mark=x] coordinates {
			(1, 1017.97286)
			(2, 1045.85791)
			(3, 977.8734)
			(4, 987.4032)
			(5, 305.530686666667)
			(6, 305.577946666667)
			(7, 297.84204)
			(8, 321.681933333334)
			(9, 156.849646)
			(10, 157.229334)
			(11, 157.530778)
			(12, 159.092124)
			(13, 180.35199)
			(14, 182.19789)
			(15, 183.61021)
			(16, 181.41781)
		};
		\addplot[color=kitgreen,mark=x] coordinates {
			(1, 683.93316)
			(2, 693.28787)
			(3, 316.8860625)
			(4, 316.6982)
			(5, 109.07319)
			(6, 110.217732)
			(7, 109.998704)
			(8, 119.05678)
			(9, 102.119487)
			(10, 102.45965)
			(11, 102.255887)
			(12, 107.41252)
			(13, 101.837897777778)
			(14, 99.6569155555556)
			(15, 102.776831111111)
			(16, 102.156475555556)
		};
		\addplot[color=kitbrown,mark=x] coordinates {
			(1, 943.40952)
			(2, 983.49588)
			(3, 287.912566666667)
			(4, 301.505426666667)
			(5, 98.6040028571428)
			(6, 98.5841185714286)
			(7, 96.7445785714286)
			(8, 103.435928571429)
			(9, 126.279816)
			(10, 126.332076)
			(11, 122.749276)
			(12, 124.80156)
			(13, 109.43448)
			(14, 108.300096)
			(15, 109.412928)
			(16, 115.371264)
		};
		\legend{AVX2 $\rightarrow$ Scalar, AVX-512 $\rightarrow$ Scalar, AVX-512 $\rightarrow$ AVX-2}
		\end{axis}
	\end{tikzpicture}
	\end{center}
	\caption{
		Break-even time for frequency changes calculated from Figure~\ref{fig:freq-reduction-overhead-results} and~\ref{fig:freq-increase-overhead-results}, assuming performance to be proportional to frequency.
		The break-even times vary with the number of active cores due to the different magnitude of the frequency change.
		The results show that a single fixed timeout as implemented by Intel CPUs can not be optimal in terms of worst-case competitiveness.
	}
	\label{fig:break-even-time}
\end{figure}

The break-even time $t_{BE}$ -- i.e., the time after which the performance increase due to increased frequencies offsets the cost to increase and decrease the frequency -- can be calculated according to the following formula:

$$
	p_{low} t_{BE} = p_{high} (t_{BE} - t_{o})
$$

In this formula, $p_{low}$ and $p_{high}$ are the performance at the lower and higher frequency, respectively, and $t_{o} = t_{o,d} + t_{o,u}$ is the total overhead for reducing ($t_{o,d}$) and increasing ($t_{o,u}$) the frequency, measured as the equivalent CPU time as in Section~\ref{sec:freqchangecost}.

If we insert the results from the last sections and calculate $t_{BE}$, we arrive at the times shown in Figure~\ref{fig:break-even-time}.
As the performance is dominated by the frequency whereas the overhead is fairly constant, the break-even time is significantly affected by the number of active cores.
For example, for a transition between AVX2 and non-AVX frequencies, the break-even time in situations with less than four active cores is approximately \SI{1000}{\mu\second} due to the low frequency swing of only \SI{100}{\mega\hertz} (see Table~\ref{tab:freqlevels}), whereas for more than eight cores frequency changes between 400 and \SI{500}{\mega\hertz} cause break-even times between 150 and \SI{190}{\mu\second}.

As Karlin et al.~\cite{karlin1994competitive} show, a fixed-timeout policy achieves optimal competitiveness -- in our case, minimal overhead when the system has to switch back to a lower frequency at the least opportunistic time -- when the timeout equals the break-even time.
In this case, the timeout before the CPU increases its frequency should therefore be based on the frequency difference to achieve good competitiveness in all cases.
Intel CPUs, however, only implement one fixed timeout for all core counts and instruction sets.
As shown in Section~\ref{sec:avxeffects}, some applications are negatively affected by the overhead of frequency changes, which shows that an improved DVFS policy with variable timeout based on the frequency difference can likely have positive impact on these applications.

\section{Exploiting Application Behavior}
\label{sec:design}

While the 2-competitive fixed-timeout policy is optimal in the worst case for unpredictable workloads, it is not when the behavior of the workload is predictable, in which case earlier decisions to increase the CPU frequency can result in higher performance.
In this work, we focus on two types of predictions about whether the system is going to use AVX-512{} in the near future.
First, the application developer has knowledge about the structure of the application and can tell the operating system when AVX-intensive parts begin and end, which can aid workloads where one process switches between AVX-intensive code and code without power-intensive instructions.
Second, the operating system can statistically determine whether a process is likely to require a reduced frequency and can change the CPU frequency during context switches in order to immediately let non-power-intensive processes profit from higher frequencies.

\subsection{Heterogeneous Applications}
\label{sec:hint_heterogeneous}

If an application consists of vectorized and non-vectorized parts and those are executed alternately -- such as the web server example in Section~\ref{sec:intro} -- the non-vectorized part is slowed down due to the frequency change caused by the vectorized part.
Often, software developers know which part of the application is vectorized and how long execution of each part takes.
In that case, assuming that a suitable hardware-software interface exists, they can notify the CPU after each vectorized code portion if the next scalar portion is likely \emph{long enough} to warrant for an early frequency increase.
The CPU could use that hint to immediately switch to a higher frequency.
Such a hint could therefore improve performance, as the existing DVFS policy of the CPU would instead needlessly keep the frequency reduced for some time.

\subsection{Classification of Tasks}
\label{sec:hint_classification}

Even if each individual application is sufficiently uniform, it is still possible that context switches between different applications cause overhead as an application is slowed down by the preceding AVX-enabled application as described in Section~\ref{sec:avxeffects}.
For most workloads, this overhead is avoidable, as scheduler time slices are usually longer than the break-even time.
During a switch from an AVX-enabled application to a non-AVX application, the scheduler should usually immediately select a higher frequency.

To trigger such frequency changes, the scheduler needs a categorization of the individual processes based on their instruction set usage and their expected frequency reduction.
To this end, we introduce the notion of a \emph{power score} which serves as a measure of the expected power consumption of the instruction mix executed by a process.
A high power score signals that the process will likely trigger significant frequency reductions.
More specifically, a power score of 1 means that the process is assumed to execute at AVX2 frequencies, whereas a power score of 2 means that the process likely causes a reduction down to AVX-512 frequency levels.

This power score could potentially be determined either via a static analysis of the application binary or via a dynamic analysis of the frequency changes at runtime.
A static analysis can detect whether an executable contains any AVX2 or AVX-512 instructions that could trigger a frequency reduction.
However, applications might contain such instructions even if they do not execute them frequently enough to significantly reduce the average frequency.
Also, functions like \texttt{memset} make use of AVX-512 instructions, but only for inputs of certain size which is hard to detect via static analysis.
Overall, a static analysis is therefore bound to be unreliable.

We expect dynamic analyses to yield a better estimate of the instruction set usage of individual processes as they are able to observe the effects of the actual execution patterns within the process.
Simply mapping the frequency level to the active process is, however, not accurate in situations with frequent context switches, because the delays mean that some of the time spent at lower frequencies is attributed to the wrong processes.
Counting the AVX2 and AVX-512 instructions executed by the active process might be sufficient to draw conclusions about the resulting frequency requirements in most cases, but recent Intel CPUs only provide performance counters for specific types of such instructions~\cite[p. 19-20f]{intelmanualvol3}.
In any case, though, more accurate statistics would be possible if the processor provided the operating system with information about whether the conditions for each frequency level were fulfilled at each point in time, for example via appropriate performance counters.
Current hardware does not provide such performance counters, either.

As a method to collect reliable information about frequency requirements and to determine the processes responsible for frequency reductions, we therefore suggest distinguishing between two cases based on the time between subsequent scheduler invocations.

If the time between subsequent scheduler invocations is significantly longer than the frequency increase delay of \SI{670}{\micro\second}, the scheduler can sample the CPU frequency level and can directly attribute the frequency to the last process, as any influence of its predecessor on the CPU frequency has ended.
To determine the CPU frequency level, we configure the performance counters to track the cycles spent at \emph{power license levels} 0, 1, and 2 which correspond to the frequency levels for non-AVX, AVX2, and AVX-512 code, respectively~\cite{optimizationmanual}.

If the time between subsequent scheduler invocations is shorter than the frequency increase delay, such an approach would risk misattributing frequency changes.
In this case, our main observation is that if the frequency is reduced during the execution of a process, then that process is most likely responsible for the change.
For short periods of execution of a process, we therefore only attribute the resulting frequency to the process in case of a frequency change during the period.
In some rare cases, however, frequency changes can occur during the execution of a process that did not trigger the change -- most likely due to delays during frequency selection as documented by Intel~\cite{optimizationmanual}.
Therefore, the power score is calculated as the moving average over all CPU frequency samples attributed to a process to reduce the impact of occasional misattribution.
The following steps are conducted to calculate the power score of the processes:
\begin{enumerate}
	\item Initially, the power score of new processes is set to 0, i.e., the system assumes that new processes will not use AVX-512 or AVX2.
	\item At each scheduler invocation, we detect the current power license level by sampling all power license level performance counters twice in a row.
		The counter that is incremented during the short time inbetween indicates the current frequency level.
	\item We compare the level during two consecutive context switches.
		If the levels match, the power license did not change.
		In this case, for short CPU bursts, the current process might not have had enough time to have an impact on the power license, so the power score is not updated.
	\item If context switches are more than \SI{1}{\milli\second} apart -- longer than the frequency delay as reasoned above -- or if the power license decreases below or increases above the current power score, however, the power score of the process is updated as the exponential moving average of such power license changes.
		Assuming $S_{t-1}$ is the old power score and $L_t$ is the new power license, the new power score is $S_t = 0.2 L_t + 0.8 S_{t-1}$.
\end{enumerate}

The resulting power score indicates the potential frequency reduction caused by the process.
The dynamic analysis of frequency changes can be combined with the results of a static analysis of the executable -- e.g., by overriding the score to be 0 if the executable does not contain AVX2 nor AVX-512 instructions -- and with manual instrumentation as described in Section~\ref{sec:hint_heterogeneous}, in which case hints from the developer override the automatically determined power score.

Note that with hyperthreading the frequency is determined by two programs. Thus, this technique only works on systems with deactivated hyperthreading and on systems which always schedule the same program on both cores, as recently suggested for the Linux kernel~\cite{corbet2019corescheduling}.
On other systems, the hardware has to be modified to provide a more reliable source of information about the energy consumption of the instructions executed by individual processes. 

\subsection{Using Hints For DVFS}

Once predictions about the instruction set use are available, the system can use this information to improve performance.
When the code running on a core -- i.e., all hardware threads in the case of a system with hardware multithreading -- indicates that no power-intensive instructions are going to be executed in the near future, for example, via the mechanisms presented in Sections~\ref{sec:hint_heterogeneous} or~\ref{sec:hint_classification}, the system can eagerly increase the frequency when it is not already at the highest level possible for the expected instructions.

Ideally, the DVFS policy should be implemented in the CPU to be able to provide quick reactions to changing instruction usage and to prevent power budget violations, so any hint about future instruction usage needs to communicated to the CPU using an appropriate software-hardware interface.
For example, the operating system or the application software could temporarily configure a different frequency change timeout depending on the type of executed code, to force earlier frequency changes or to prevent any changes.

\subsubsection{Viability on Current CPUs}

Current hardware does not provide any such interface.
It does, however, provide a mechanism to manually set the CPU frequency, which can be used to implement a wide range of DVFS policies in software.
For the dim silicon scenario described in this paper, the limitations of the hardware prevent both practical software-based implementations of DVFS policies as well as limited implementations to estimate the performance of hardware-based implementations.

Any practical implementation of a DVFS solution for AVX-512 or AVX2 code is prevented both by the inability to detect problematic AVX-512 or AVX2 code as well as by the delay of manual frequency changes.
First, conservative detection of problematic code is necessary so that the OS knows when frequency reductions are required.
Our approach in Section~\ref{sec:hint_classification} is not usable as it only results in approximate long-term classification of applications.
In contrast, conservative short-term estimation based on register set usage can detect any access to 512-bit and 256-bit vector registers but will often select lower frequencies than necessary as we show in our evaluation in Section~\ref{sec:categorization-eval}, leading to reduced performance.
Second, software-based DVFS policy implementations require the ability to change the frequency at a precise point in time, yet current CPUs delay frequency changes significantly.
As described by Hackenberg et al.~\cite{hackenberg2015energy}, the frequency selection logic of Intel CPUs starting with the Haswell microarchitecture only allows frequency changes once every \SI{500}{\micro\second}, so any frequency change request is delayed until the end of the next such \SI{500}{\micro\second} window.
The immediate throttling of AVX-512 instructions~\cite{downs20gathering}, however, shows that immediate power reduction is necessary for stability, so such delays are inacceptable.

These limitations not only prevent practical software solutions but unfortunately also prevent the construction of a prototype based on existing hardware to evaluate the performance of hardware implementations.
Such a prototype would not necessarily have to be able to ensure system stability, but would have to trigger frequency changes in a way that results in equal performance compared to a complete implementation.
As from the point of view of the OS the frequency change delay often appears to be random with an even distribution, a na\"ive approach might assume that the average delay of frequency increases cancels out the average delay of frequency reductions.
However, for short sections of AVX- or non-AVX code, both frequency increase and decrease might occur within the same \SI{500}{\micro\second} window, in which our experiments showed that no frequency change occurs.

In this paper, we suggest improved DVFS policies as a method to reduce the overhead caused by AVX2 and AVX-512.
As we cannot use existing hardware to conduct a performance evaluation, we are limited to demonstrating the performance impact through simulations and microbenchmarks as shown in Section~\ref{sec:dvfs-policy-eval}.

\section{Evaluation}
\label{sec:evaluation}

As described in the last section, this paper proposes using hints from the application or the operating system to provide improved frequency scaling.
Our approach consists of two main pieces, namely the classification of the processes -- or, alternatively, hints from the application developer -- and a modified DVFS algorithm that takes those hints into account.
For existing processors, it is impossible to build a complete implementation of this design, as the existing DVFS policy implemented by the CPU cannot be extended as required.
Deactivating all AVX-induced frequency changes and completely reimplementing the policy in software is impossible due to the latency of software-triggered frequency changes which can be as long as 500\,$\mu$s.
Our evaluation is therefore limited to qualitatively showing that the individual components are functional and that application-directed DVFS can have an advantage over the existing policy.

\subsection{Categorization of Processes}
\label{sec:categorization-eval}

\begin{table*}[t]
	\begin{center}
		\begin{tabular}{|r|c|c|c|}
			\hline
			& Predominant & & Average \\
			Scenario & Freq. Level & AVX Score & \texttt{AVX512\_elapsed\_ms} \\
			\hline
			\hline
			x265 (AVX) & 0 & 0.356 & N/A \\
			\hline
			x265 (AVX2) & 1 & 1.005 & N/A \\
			\hline
			x265 (AVX-512) & 2 & 1.899 & 103.4\,ms \\
			\hline
			x265 (AVX-512) & 2 & 1.827 & 181.3\,ms \\
			+ pts-apache & 0 & 0.428 & N/A \\
			\hline
			x265 (AVX-512) & 2 & 1.679 & 86.0\,ms \\
			+ parsec-swaptions & 0 & 0.099 & N/A \\
			\hline
			\hline
			512-bit FMA & 2 & 1.821 & 0.10\,ms \\
			\hline
			512-bit add & 1 & 0.917 & 0.10\,ms \\
			\hline
			256-bit FMA & 1 & 0.934 & N/A \\
			\hline
			256-bit add & 0 & 0 & N/A \\
			\hline

		\end{tabular}
	\end{center}
	\caption{
		Estimated AVX scores for different scenarios and a comparison to the mechanism found in the Linux kernel to track AVX-512 usage.
		The first three rows show the score for an isolated instance of x265 using different instruction sets.
		The next two rows show the scores in scenarios with two different applications running concurrently on the same set of cores, to show that the score is estimated correctly on a per-process basis even if one process affects the frequency of another.
		The remaining rows show how our approach is able to distinguish between the three frequency levels of the CPU, whereas the stock Linux kernel is only able to track AVX-512 register usage.
	}
	\label{tab:categorization-eval}
\end{table*}

The main goal of the process classification mechanism described in Section~\ref{sec:hint_classification} is to be able to detect the required power license of individual processes even if they are running in a heterogeneous multi-process workload where the effects of one process on the CPU frequency might shadow the effects of another process.
To show that the mechanism fulfills this goal, we constructed a prototype based on Linux 5.2.
We modified the kernel's completely fair scheduler (CFS) and inserted the power license detection code in the main scheduler function \texttt{\_\_schedule()}.
Our implementation uses the Linux perf framework to read the power license performance counters.

We let our prototype estimate the power score of the x265 video encoder using different instruction sets running in isolation.
To show that our prototype is able to correctly distinguish between different processes executing on the same system and is able to attribute frequency changes to the correct process, we also executed the Apache benchmark from the Phoronix Test Suite as well as the swaptions benchmark from Parsec in parallel with x265.
The two applications were configured to share the same set of cores without any restrictions to scheduling.
Note that we specifically selected an interactive benchmark as well as a batch workload, to show that the classification works with both.
Table~\ref{tab:categorization-eval} shows both the expected power score for the applications -- we expected our prototype to classify x265 according to the instruction set used, and neither of the other two benchmarks used significant amounts of vector instructions -- as well as the estimated power score from our prototype, averaged over the runtime of the application.

The first three rows show that x265 was correctly classified in all cases, except for some uncertainty if neither AVX2 nor AVX-512 instructions were used. 
The next two table rows then show the results for the mixed scenarios.
In both cases, our prototype can correctly identify x265 as the process responsible for the frequency reduction.
For x265 executed alone, we compared the performance of our prototype to a stock Linux kernel and were not able to measure any statistically significant performance overhead.

We compare our approach to the state-of-the-art technique available in the Linux kernel.
Linux provides the time elapsed since the last use of AVX-512{} as part of the \emph{arch\_status} file in the proc file system~\cite{linuxproc}.
The time since the last use of AVX-512 is calculated by checking the state of the FPU registers at each context switch. 
Like our approach, this mechanism is able to detect AVX-512 usage in the benchmarks described above as shown in the upper half of Table~\ref{tab:categorization-eval}.

The approach found in the Linux kernel has a significant drawback, though, as the use of specific FPU registers is only loosely connected to the resulting frequency change.
For example, a dense sequence of multiplication instructions on 512-bit vector registers causes the CPU to transition to the lowest frequency, whereas other instructions only trigger the intermediate \enquote{AVX2} frequency.
Therefore, in a workload consisting of processes showing the former behavior as well as processes of the latter type, the time since the last 512-bit register usage cannot be used to identify the processes responsible for a frequency reduction.
We demonstrate this effect by executing a sequence of 512-bit and 256-bit multiplications and additions both with our approach and on an unmodified Linux 5.5 kernel.
The results shown in the lower half of Table~\ref{tab:categorization-eval} show that our prototype is correctly able to detect the three different frequency levels caused by different types of instructions, whereas the stock Linux kernel is only able to detect whether 512-bit registers are used.

To show that the problem also affects real-world workloads, we execute an web server benchmark using nginx and OpenSSL similar to the one described in Section~\ref{sec:avxeffects} and measure the average time since the last AVX-512 usage as determined by the Linux 5.5 kernel on a system running Fedora 30.
We let the nginx web server serve a static file with compression at runtime and use OpenSSL compiled with either AVX2 and AVX-512 instruction support for TLS encryption.
As shown above, the web server provides significantly higher performance when using AVX2 instructions due to the resulting higher frequencies.
Even in this case the system uses 512-bit registers, though, as the C library provides AVX-512 variants of \texttt{memset()}, \texttt{memmove()}, and \texttt{memcpy()}.
Therefore, the stock Linux kernel detects AVX-512 usage in both cases, with similar reported average time since the last usage of 512-bit registers.
Note that the implementation tests whether registers are in use only during context switches.
Different scheduling causes large variation in the resulting values, making a quantitative comparison for such experiments difficult.

\subsection{Potential of Eager Frequency Changes}
\label{sec:dvfs-policy-eval}

Once it is known which parts of the system use power-intensive instructions -- either using manual annotation as described in Section~\ref{sec:hint_heterogeneous} or via automatic detection as described in the previous section -- this information can be used to optimize performance.
Whereas other approaches perform core specialization to separate AVX-512 code from non-AVX code~\cite{li2019corescheduling,gottschlag19sfma}, we, as described in Section~\ref{sec:design}, suggest that improved DVFS policies can also significantly reduce the overhead caused by AVX-512 instructions and similarly power-intensive instruction sets.
In particular, we suggest that the delay for frequency increases as implemented by recent Intel CPUs is unnecessary if the system can predict that the software executed in the near future does not require power-intensive instructions.


\subsubsection{Methodology}
\label{sec:eval-methodology}

\tikzstyle{block1} = [rectangle, draw, text width=5em, text centered, rounded corners, minimum height=3em]
\tikzstyle{block2} = [rectangle, draw, text width=5em, text centered, minimum height=3em]
\tikzstyle{line} = [draw, -latex']

\begin{figure}[t]
	\begin{center}
	\begin{tikzpicture}[node distance = 2.3cm, auto]
		\node [block1] (trace) {application trace};
		\node [right of=trace] (dummy) {};
		\node [block1, right of=dummy] (costs) {frequency change costs (Section~\ref{sec:freqchangecost})};
		\node [block2, below of=dummy] (sim) {simple model-based simulator};
		\node [block2, right of=sim] (policy) {policy};
		\node [block1, below of=sim] (runtime) {runtime};

		\draw [->] (trace) -| (sim.120);
		\draw [->] (costs) -| (sim.60);
		\draw [<->] (sim) -- (policy);
		\draw [->] (sim) -- (runtime);
	\end{tikzpicture}
	\end{center}
	\caption{
		Experimental setup to estimate the performance resulting from different DVFS policies.
		Our simulator implements a very simple performance model assuming fixed frequency change costs and performance proportional to the CPU frequency.
		We instrumented the nginx web server and used the resulting trace of AVX- and non-AVX periods to determine the runtime with different DVFS policies.
	}
	\label{fig:eval-methodology}
\end{figure}
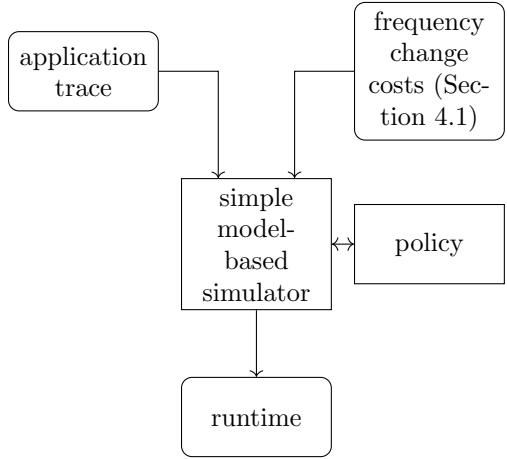

The most direct method to show the potential of improved DVFS policies would be to compare the performance of a benchmarked application when using a fixed-timeout policy such as the one implemented by the processor to the same benchmark instrumented to change the processor frequency at points in the program selected by the developer.
However, recent Intel CPUs delay frequency change requests by up to \SI{500}{\micro\second}\cite{hackenberg2015energy}, making it impossible to precisely specify the points in the program at which frequency changes occur.
Therefore, our evaluation relies on simulation of different DVFS policies based on a trace generated while running a web server benchmark (Section~\ref{sec:eval-simulation}) and uses a microbenchmark to demonstrate the potential performance impact of a single eager frequency change (Section~\ref{sec:eval-microbenchmark}) and to check the accuracy of the simulation.

The following experiments were conducted on a system with an Intel Core i9-7940X processor, with the simulation configured to match this system.
This processor was selected because, as it is designed for overclocking, it allows configuration of the \emph{AVX offsets} which specify the frequency reduction caused by AVX2 and AVX-512 instructions. 
For tests to determine the baseline performance of the system, we configured the offsets to match the frequencies reported in news articles~\cite{spille2017skylake}, where the base frequency of the processor is reported to be \SI{3.1}{\giga\hertz} and the frequencies for AVX2 and AVX-512 code are \SI{2.7}{\giga\hertz} and \SI{2.4}{\giga\hertz}, respectively, providing similar frequency ratios compared to server processors.
Note that no authoritative information about AVX2 and AVX-512 frequencies is found in official Intel documentation and that mainboards such as ours frequently provide non-default AVX offsets.

With minimal AVX offsets, the AVX2 and AVX-512 frequencies are both \SI{3.0}{\giga\hertz}, so the AVX frequency reduction cannot be disabled completely.
We discuss the effect of this minimum frequency change where applicable below.
All experiments were executed with Turbo Boost disabled and with C-states limited to C1 in order to reduce variance in the measurement results.

\subsubsection{Web Server Simulation}
\label{sec:eval-simulation}

For our simulation experiments, the workload used is the nginx web server example from Section~\ref{sec:avxeffects}.
We configure the web server to serve a single static file using gzip compression and we encrypt HTTP requests and replies using the OpenSSL library.
The library is configured to vectorize encryption and decryption using AVX-512 instructions, which in other experiments has resulted in a 10\% slowdown.
We instrument the web server to record the times when the OpenSSL functions for encryption and decryption are called and when they return.
When generating the log of the OpenSSL function calls, we execute the benchmark with minimal AVX offsets.
Although the resulting frequencies would not be stable and would result in frequent system crashes with all cores utilized, this setup yields more representative timing input for the simulator, as the simulator itself is supposed to slow down the AVX-512 portions of the simulated workload.
To ensure system stability and to simplify simulation, the web server is only executed on a single core.
We do not expect individual web server threads to behave significantly different when additional web server threads are placed on the other cores of the system.

The resulting application trace contains a list of periods where the system is assumed to execute only AVX-512 code (the function calls into OpenSSL) alternating with periods where the system is assumed not to execute any AVX-512 or AVX2 instructions.
We feed this trace into a simple model-based simulator as shown in Figure~\ref{fig:eval-methodology} to estimate the application runtime resulting from different DVFS policies.
The simulator applies a DVFS policy to the trace and dilates the time during periods where the CPU would be executing at a lower frequency.
During the simulation, to get results more representative for a server scenario, we assume that most of the cores are active and assume a corresponding large frequency reduction whenever AVX-512 code is executed.

We implement fixed-timeout policies with the timeout used by Intel processors as well as with a timeout of \SI{180}{\micro\second} which was shown to be more competitive in Section~\ref{sec:break-even-time}.
As an example for a policy based on developer input, we also implement a policy which only increases the frequency when the last packet of an HTTP request was received and decrypted, which we identify by the return value of the corresponding OpenSSL function call~\footnote{A more generic and robust implementation would be to instrument the HTTP request parsing logic to increase the frequency whenever the end of a HTTP request is detected. Our implementation suffices to show that the approach is generally possible.}.
After this call, the web server processes the request and takes a significant amount of time before any further AVX-512 code is executed when the HTTP reply is sent, so at this point eager frequency changes are most likely to be beneficial for application performance.
For all the policies, the simulator assumes a performance impact of \SI{16}{\micro\second} per frequency change, similar to the values determined experimentally in Section~\ref{sec:freqchangecost}. 

The simulation result shows that a lower timeout than what is used by Intel CPUs results in a 2.9\% higher performance in the simulated scenario.
With a lower timeout, the policy can exploit shorter non-AVX program phases and wastes less time at lower frequencies throughout the program.
The resulting performance improvement outweighs the (simulated) overhead of the larger number of frequency changes.
Even though the difference is small, the result shows that the timeout does have a measurable impact on application performance.

The developer-directed DVFS policy performed even better, with a 3.9\% performance improvement compared to the policy implemented by Intel CPUs, as the policy was able to completely mitigate overhead due to low CPU frequencies during the longest non-AVX phases of the program.
While this improvement might seem minor, it covers most of the 5.7\% overhead caused by AVX-512 for this workload as shown in Figure~\ref{fig:overhead-without-ht}.
Workloads with more frequent AVX-512 phases might benefit more from improved policies.
In addition, the policy did not increase the frequency during some other non-AVX phases where a frequency change would have been beneficial, showing that a carefully optimized prototype might achieve higher performance.

\subsection{Maximum Potential per Frequency Change}
\label{sec:eval-microbenchmark}

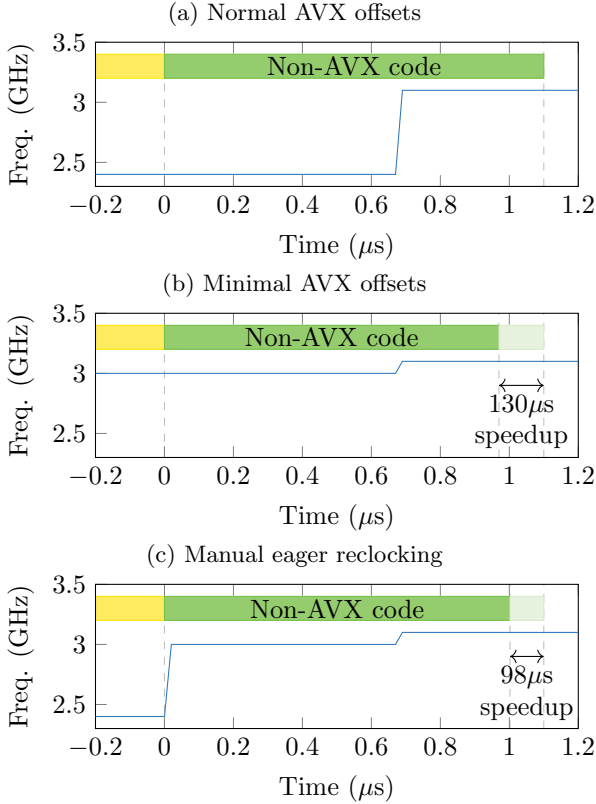
\begin{figure}[t]
	\begin{subfigure}[c]{0.45\textwidth}
		\begin{center}
		\subcaption{
			Normal AVX offsets
		}
		\label{fig:frequency-change-normal-offsets}
		\vskip -3mm
		\begin{tikzpicture}
			\begin{axis}[
				xlabel=Time ($\mu$s),
				ylabel=Freq. (GHz),
				height=3.5cm,
				width=8cm,
				ymin=2.3,
				ymax=3.5,
				xmin=-0.2,
				xmax=1.2,
				ytick pos=left,
				axis background/.style={preaction={path picture={
					\draw [gray!50,sharp plot,dashed] (axis cs:0,2.3) -- (axis cs:0,3.6);
					\draw [gray!50,sharp plot,dashed] (axis cs:1.1,2.3) -- (axis cs:1.1,3.6);
					\draw[color=kityellow,fill=kityellow!75] (axis cs:-0.2,3.2) rectangle (axis cs:0,3.4);
					\draw[color=kitgreen,fill=kitgreen!75] (axis cs:0,3.2) rectangle (axis cs:1.1,3.4) node[pos=.5, color=black] {Non-AVX code};
				}}},
			]
			\addplot[color=kitblue] coordinates {
				(-0.2, 2.4)
				(0.670, 2.4)
				(0.690, 3.1)
				(1.2, 3.1)
			};
			\end{axis}
		\end{tikzpicture}
		\end{center}
	\end{subfigure}
	\begin{subfigure}[c]{0.45\textwidth}
		\begin{center}
		\subcaption{
			Minimal AVX offsets
		}
		\label{fig:frequency-change-minimal-offsets}
		\vskip -3mm
		\begin{tikzpicture}
			\begin{axis}[
				xlabel=Time ($\mu$s),
				ylabel=Freq. (GHz),
				height=3.5cm,
				width=8cm,
				ymin=2.3,
				ymax=3.5,
				xmin=-0.2,
				xmax=1.2,
				ytick pos=left,
				axis background/.style={preaction={path picture={
					\draw [gray!50,sharp plot,dashed] (axis cs:0,2.3) -- (axis cs:0,3.6);
					\draw [gray!50,sharp plot,dashed] (axis cs:0.97,2.3) -- (axis cs:0.97,3.6);
					\draw [gray!50,sharp plot,dashed] (axis cs:1.1,2.3) -- (axis cs:1.1,3.6);
					\draw[color=kityellow,fill=kityellow!75] (axis cs:-0.2,3.2) rectangle (axis cs:0,3.4);
					\draw[color=kitgreen,fill=kitgreen!75] (axis cs:0,3.2) rectangle (axis cs:0.97,3.4) node[pos=.5, color=black] {Non-AVX code};
					\draw[color=kitgreen!25,fill=kitgreen!15] (axis cs:0.97,3.2) rectangle (axis cs:1.1,3.4);
					\draw[<->] (axis cs:0.97,2.9) -- node[below, align=center] {130$\mu$s\\speedup} (axis cs:1.1,2.9);
				}}},
			]
			\addplot[color=kitblue] coordinates {
				(-0.2, 3.0)
				(0.670, 3.0)
				(0.690, 3.1)
				(1.2, 3.1)
			};
			\end{axis}
		\end{tikzpicture}
		\end{center}
	\end{subfigure}
	\begin{subfigure}[c]{0.45\textwidth}
		\begin{center}
		\subcaption{
			Manual eager reclocking
		}
		\label{fig:frequency-change-manual-reclocking}
		\vskip -3mm
		\begin{tikzpicture}
			\begin{axis}[
				xlabel=Time ($\mu$s),
				ylabel=Freq. (GHz),
				height=3.5cm,
				width=8cm,
				ymin=2.3,
				ymax=3.5,
				xmin=-0.2,
				xmax=1.2,
				ytick pos=left,
				axis background/.style={preaction={path picture={
					\draw [gray!50,sharp plot,dashed] (axis cs:0,2.3) -- (axis cs:0,3.6);
					\draw [gray!50,sharp plot,dashed] (axis cs:1.002,2.3) -- (axis cs:1.002,3.6);
					\draw [gray!50,sharp plot,dashed] (axis cs:1.1,2.3) -- (axis cs:1.1,3.6);
					\draw[color=kityellow,fill=kityellow!75] (axis cs:-0.2,3.2) rectangle (axis cs:0,3.4);
					\draw[color=kitgreen,fill=kitgreen!75] (axis cs:0,3.2) rectangle (axis cs:1.002,3.4) node[pos=.5, color=black] {Non-AVX code};
					\draw[color=kitgreen!25,fill=kitgreen!15] (axis cs:1.002,3.2) rectangle (axis cs:1.1,3.4);
					\draw[<->] (axis cs:1.002,2.9) -- node[below, align=center] {98$\mu$s\\speedup} (axis cs:1.1,2.9);
				}}},
			]
			\addplot[color=kitblue] coordinates {
				(-0.2, 2.4)
				(0, 2.4)
				(0.02, 3.0)
				(0.670, 3.0)
				(0.690, 3.1)
				(1.2, 3.1)
			};
			\end{axis}
		\end{tikzpicture}
		\end{center}
	\end{subfigure}
	\caption{
		To determine the potential performance improvement for a single frequency change, we execute a fixed amount of non-AVX code directly following some AVX-512 code.
		We compare the time required at default AVX offsets (a) to the time required at minimal AVX offsets (b) as well as with eager frequency changes simulated by a manually inserted frequency change  at the beginning of the non-AVX code (c).
	}
	\label{fig:frequency-change-potential-eval}
\end{figure}

When looking at a single developer-directed eager frequency change, the simulation resulted in a CPU time saving of \SI{195}{\micro\second} for sufficiently long stretches of non-AVX code compared to a fixed timeout of \SI{670}{\micro\second} as implemented by current Intel CPUs, as the CPU was operating 30\% faster during this time.
To show that the assumptions made in our simulator yield realistic results, we validate this value against measurements based on a simple microbenchmark.
The microbenchmark first executes a series of AVX-512 instructions and then executes a fixed amount of non-AVX instructions.
The number of instructions is chosen so that they take longer than the frequency change timeout implemented by the CPU.
We measure the time required for the code section to determine the impact of the frequency change caused by the preceding AVX-512 code in different configurations.
All experiments are repeated 1000 times.

First, to measure the overall impact of such frequency changes on the CPU, we compare the average time at default frequencies (Figure~\ref{fig:frequency-change-normal-offsets}) with the average time with minimal AVX offsets (Figure~\ref{fig:frequency-change-minimal-offsets}).
Our experiment shows that with minimal frequency changes the code executes \SI{130}{\micro\second} faster.
In this configuration the AVX-512 code still reduces the CPU frequency by \SI{100}{\mega\hertz} as described in Section~\ref{sec:eval-methodology} and the measured runtime still includes the overhead of the corresponding frequency change which needs to be taken into account when comparing the values with the model used for our simulation.

Second, we manually insert frequency changes into our prototype so that the frequency is reduced when the AVX-512 code starts and is immediately increased when the non-AVX code starts.
Note that, as described in Section~\ref{sec:eval-methodology}, frequency changes are applied with a random delay of up to \SI{500}{\micro\second}.
Therefore, for this experiment, we do not take the average time but instead take the 5th percentile as this value represents the situation when an optimized DVFS policy implementation almost immediately triggers frequency changes.
In this experiment, we measure a runtime for the non-AVX code which is \SI{32}{\micro\second} slower than the result with minimum frequency changes, but \SI{98}{\micro\second} faster than regular frequency changes (Figure~\ref{fig:frequency-change-manual-reclocking}).
The performance is slightly lower than in the experiment with minimal AVX offsets because the benchmark triggers not one but two frequency changes -- one by the hardware due to the \SI{100}{\mega\hertz} reduction described above, and one manual frequency change to simulate the DVFS policy.
Apart from this overhead and minor overhead due to the additional system calls, the runtime mostly matches the optimal case, which supports our model that eager frequency changes can mitigate most of the overhead caused by AVX instructions.

However, the absolute runtime differences are lower than determined by the simulation.
As described above, two potential reasons for the deviation are the larger number of frequency changes as well as some remaining frequency reduction.
As shown in Figure~\ref{fig:freq-increase-overhead-results}, the additional frequency change costs approximately \SI{10}{\micro\second}, and an expected 3\% performance overhead due to the \SI{100}{\mega\hertz} frequency difference costs another \SI{20}{\micro\second}.
While the measured results mostly match our model when taking these effects into account, further analysis of the CPU behavior has to be conducted to provide a better quantitative model of the performance in similar situations.

\section{Discussion}
\label{sec:discussion}

In this paper, we showed that the fixed timeout policy implemented by recent Intel CPUs for AVX frequencies yields less-than-optimal average processor frequencies for heterogeneous workloads.
We also argue that better timeouts and developer-directed frequency changes can improve performance.
Even though our evaluation lacks experiments to directly demonstrate the effects on real-world workloads, the estimate generated by our simulation shows that it is highly likely that such a performance improvement is to be expected.
This basic result opens up a number of further research questions which we will discuss in the following sections.

\subsection{Hardware Interfaces}

In Section~\ref{sec:eval-methodology}, we show why the frequency change delays on current Intel CPUs prevent constructing a full prototype demonstrating our approach.
Even if frequency changes were triggered instantly, though, a software-only DVFS policy implementation would not be viable for two reasons:
First, the CPU would still need to be able to autonomously reduce power consumption when executing AVX-512 instructions to ensure system stability, for example, by reducing the frequency or applying other forms of throttling.
Second, not all applications in the system would be modified to make use of developer-directed frequency scaling, making a hardware fallback necessary.

If the DVFS policy is implemented in hardware, a software-hardware interface is required to influence policy decisions.
We propose the combination of two such interfaces:

\begin{enumerate}
\item \textbf{Configurable frequency change delay:}
As we show, the problem of AVX-induced frequency changes is similar to the dynamic power management problem, and the main decision is whether to immediately increase the frequency when possible or whether to wait or not increase the frequency at all.
While it would be possible to tell the CPU to immediately increase the frequency after the next section of AVX code, we expect such an interface not to be viable in many situations, because the boundaries of AVX-intensive program execution phases are not well defined and variations in the program's control flow might cause unnecessary frequency changes.
Instead, we suggest an interface to manually set a different frequency change timeout for individual parts of the program -- i.e., until the application manually reverts the change or sets a different timeout -- to allow applications to enable eager frequency changes in certain situations.

\item \textbf{Forced immediate frequency change:}
In addition, the CPU should provide an interface to immediately increase the frequency to the maximum frequency for use by the operating system to increase the frequency during context switches when it is known that the next task is unlikely to use AVX-512 or AVX2.
\end{enumerate}

Further work has to be conducted to test whether these interfaces are sufficiently flexible to implement a wide range of DVFS policies in software.

\subsection{Hardware Multithreading}

One significant limitation of our work is that all our experiments were conducted with a system in mind that does not use hardware multithreading.
On a system with hardware multithreading, the CPU frequency has to be reduced when either of the threads executes AVX instructions, thereby limiting the potential performance advantage of developer-directed approaches as it is hard to predict when another completely unrelated hyperthread will affect the frequency.
Also, as shown in Section~\ref{sec:avxeffects} on systems with hyperthreading many additional types of workloads experience slowdown due to frequency reductions.
Despite the differences, improved DVFS policies might be viable and their effectiveness might even be amplified as more code is affected by frequency reductions.
More research should be conducted to create a statistical model of the CPU frequency selection in systems with hardware multithreading and to develop suitable DVFS policies.

\subsubsection{AVX Overhead Profiling}

In our controlled experiment, we used a benchmark that had a clearly defined performance metric.
In general it is, however, not always clear whether the overhead caused by AVX-512 is large enough to warrant the usage of techniques to reduce it and it is not always clear whether these techniques are successful.
In particular when techniques have the potential to cause additional overhead -- for example, due to increased numbers of frequency changes -- it would be beneficial to be able to profile a system to estimate the impact of AVX-512 on performance.
The result of such a profiler could also be used to implement close-loop policies.
For example, the system could repeatedly try out different DVFS policies depending on the resulting performance change.

The performance counters on current CPUs, however, cannot be used to construct such a profiler, as they can only be used to count cycles spent at reduced frequencies but do not provide sufficient information about how long the reduced frequencies are actually required.
In particular, the performance monitoring units of these CPUs can not be used to detect any executed AVX2 and AVX-512 instructions as they can only count floating point instructions.
Instead, we envision an approach which periodically samples the frequency of the system, pauses the system to let the CPU switch back to the highest possible frequency, and then checks whether the system will immediately switch back to a lower frequency when the workload is continued.
The latter check determines whether a frequency reduction is required due to ongoing AVX code or whether the reduction represents avoidable overhead.
Further experiments have to determine the accuracy of such an approach, and further work has to be conducted to show whether modified hardware-software interfaces can provide a more accurate profiling mechanism with lower CPU time overhead.

\section{Related Work}
\label{sec:relwork}

This paper presents improved DVFS policies as a method to reduce the overhead of the frequency reduction caused by AVX and AVX-512 instructions on recent Intel CPUs.
Other approaches to this and similar problems have used core specialization or have modified the application to reduce the impact of varying power consumption and of frequent frequency changes.

\subsection{Core Specialization}

Another method to limit the performance impact of AVX and AVX-512 code on unrelated non-AVX code is to place AVX and non-AVX parts of the workload on separate sets of cores.
As performance problems occur when non-AVX code is executed on the same core following AVX code which reduced the frequency, specialization of cores can prevent such overhead.
Approaches for core specialization either targeted heterogeneous programs consisting of AVX and non-AVX code within one process~\cite{gottschlag19sfma} or targeted workloads consisting of AVX and non-AVX processes~\cite{li2019corescheduling}.
The former detects the usage of AVX instructions either by instrumentation inserted by the developer or by reconfiguring the CPU to trigger exceptions when executing AVX instructions~\cite{gottschlag19sfma,gottschlag2020automatic}.
Based on this information, individual threads are migrated between cores to concentrate the AVX part of the program on as few cores as possible.
The latter technique which is targeted at multi-process workloads instead relies on heuristics to identify processes using AVX-512 instructions and modifies the scheduler to prevent scheduling an AVX-512 and a non-AVX task on hardware threads of the same core at the same time~\cite{li2019corescheduling}.
This approach currently uses the Linux \texttt{arch\_status} interface which only gives a rough estimate of AVX-512 usage.
In this paper, we present a method to identify applications which cause frequency reductions with higher accuracy.

Note that all these approaches can cause significant performance overhead themselves.
Task migrations can increase cache miss rates, and restricting scheduling of different processes on the same core at the same time can cause significant overhead with some workloads~\cite{corbet2019corescheduling}.
We present a technique which might provide advantages in situations where other approaches cause too much overhead.

The fact that co-scheduling applications on the hardware threads of a single core can cause varying overhead depending on the type of the applications has been observed by other works before and many scheduling techniques have been developed to improve the performance of SMT systems.
For example, existing approaches use sampling-based techniques~\cite{snavely2000symbiotic}, cache conflict detection~\cite{settle2004architectural}, or performance counters~\cite{el2006compatible,mcgregor2005scheduling} to determine whether two tasks are suited for parallel scheduling on the same physical core.
We describe a similar approach which uses performance counters to identify tasks requiring execution at reduced frequency and which can likely be used for improved co-scheduling of AVX-512 applications as described above.

\subsection{Profile-Guided Software Modifications}

The approach in this paper is designed either for applications which are only available in binary form or which can benefit from AVX2 and AVX-512 instructions.
If a program only makes use of such instructions in very short execution phases, those parts could alternatively be rewritten to use instructions with lower power consumption.

Kumar et al.~\cite{kumar2014efficient} use such an approach to improve the efficiency of power-gating the processor's SIMD unit.
In this scenario, devectorizing parts of the program reduces the speedup caused by SIMD instructions, but reduces the power-gating overhead.
The authors use a profiler to determine the SIMD instruction usage in individual parts of the program.
As static recompilation based on this information is problematic as the profiling results are only accurate for specific input data, the authors integrate the profiler into a system which uses dynamic translation at runtime to devectorize those parts which only rarely use SIMD instructions.
Such an approach could likely be applied to AVX-512 to improve average CPU frequencies, although hardware modifications would be required -- current CPUs can only count floating-point AVX-512 instructions, but not integer operations~\cite[p. 19.20f]{intelmanualvol3}.
Even with such hardware changes, it is not possible to use the approach with existing ahead-of-time compilers, though.
In our work, we explore techniques usable within the existing software environment.

Roy et al.~\cite{roy2009framework}, instead, suggest a similar technique that uses information from dynamic profiling to insert static power management code into an application at compile time.
Their approach inserts instructions for power gating of parts of the processor in order to save energy.
A similar approach, however, could potentially be used to let the application guide frequency selection decisions of the processor.

\section{Conclusion}
\label{sec:conclusion}

Modern Intel CPUs reduce their frequency whenever power-intensive AVX2 or AVX-512 instructions are executed to prevent violating power limits.
The frequency is only increased again after a fixed timeout has elapsed, in order to prevent excessive numbers of frequency changes.
This behavior reduces the performance for heterogeneous workloads where code sections with and without such AVX instructions alternate, as parts of the latter are executed at a lower frequency than necessary.

We show the similarity between this behavior and mechanisms from dynamic power management.
We show that the constant delay before increasing the frequency is not optimal in terms of worst-case competitiveness and show how the delay should depend on the magnitude of the frequency change.
We also sketch how information from the OS or the developer can be used to inform the CPU about future system behavior so that the CPU can implement more efficient DVFS policies.
Although we do not have a complete implementation due to constraints of the hardware, we show that it is possible to reliably determine whether an application will cause frequency changes and we show that eager frequency changes based on such information about the workload can improve performance.

\subsection{Future Work}
\label{sec:futurework}

Although we show that an oracle-style DVFS policy can improve performance, it remains to be seen whether other approaches from the area of dynamic power management can be applied as well.
In particular, some shutdown strategies achieve lower power consumption compared to the simple fixed-timeout policy even without application-level knowledge.

In addition, due to hardware constraints, we do not present any complete implementation of our approach.
We plan to construct a testbed for other DVFS policies and to use it to evaluate different hardware-software interfaces which would allow input from the operating system or from applications to affect hardware-controlled frequency scaling.

\appendix

\bibliographystyle{abbrv}
\bibliography{techreport}

\end{document}